\documentclass[aps,prl,floatfix,reprint,showpacs,superscriptaddress,groupedaddress,twocolumn]{revtex4-1}
\usepackage{latexsym}
\usepackage{graphicx}
\usepackage{rotating}
\usepackage[normalem]{ulem}
\usepackage{hyperref}
\usepackage{amsmath,amssymb,amsfonts}
\usepackage{pifont}
\usepackage{bm}
\usepackage{color}
\usepackage{soul}  
\usepackage{wasysym}

\usepackage{color}

\newcommand{\lyxmathsym}[1]{\ifmmode\begingroup\def\b@ld{bold}
  \text{\ifx\math@version\b@ld\bfseries\fi#1}\endgroup\else#1\fi}

\usepackage{amsfonts}
\usepackage{mathrsfs}

\hypersetup{colorlinks=true,urlcolor=blue,linkcolor=blue}

\begin{document}

\title{Interplay between band structure and Hund's correlation to increase T$_{c}$ in FeSe}
\author{Swagata Acharya$^{1}$}
\affiliation{ King's College London, Theory and Simulation of Condensed Matter,
              The Strand, WC2R 2LS London, UK}
\affiliation{ National Renewable Energy Laboratories, Golden, CO 80401}
\email{swagata.acharya@kcl.ac.uk}
\author{Dimitar Pashov$^{1}$}
\affiliation{ King's College London, Theory and Simulation of Condensed Matter,
              The Strand, WC2R 2LS London, UK}
\affiliation{ National Renewable Energy Laboratories, Golden, CO 80401}
\author{Francois Jamet$^{1}$}
\affiliation{ King's College London, Theory and Simulation of Condensed Matter,
	The Strand, WC2R 2LS London, UK}
\affiliation{ National Renewable Energy Laboratories, Golden, CO 80401}
\author{Mark van Schilfgaarde$^{1,2}$}
\affiliation{ King's College London, Theory and Simulation of Condensed Matter,
              The Strand, WC2R 2LS London, UK}
\affiliation{ National Renewable Energy Laboratories, Golden, CO 80401}


\begin{abstract}

FeSe is classed as a Hund's metal, with a multiplicity of $d$ bands near the Fermi level.  Correlations in Hund's metals
mostly originate from the exchange parameter \emph{J}, which can drive a strong orbital selectivity in the correlations.
The Fe-chalcogens are the most strongly correlated of the Fe-based superconductors, with $d_{xy}$ the most correlated
orbital.  Yet little is understood whether and how such correlations directly affect the superconducting instability in
Hund's systems.

By applying a recently developed high-fidelity \emph{ab initio} theory, we show explicitly the connections between
correlations in $d_{xy}$ and the superconducting critical temperature $T_{c}$.  Starting from the \emph{ab initio}
results as a reference, we consider various kinds of excursions in parameter space around the reference to determine
what controls $T_{c}$.  We show small excursions in $J$ can cause colossal changes in $T_{c}$.  Additionally we consider
changes in hopping by varying the Fe-Se bond length in bulk, in the free standing monolayer M-FeSe, and M-FeSe
on a SrTiO$_{3}$ substrate (M-FeSe/STO). The twin conditions of proximity of the $d_{xy}$ state to the Fermi energy, and
the strength of $J$ emerge as the primary criteria for incoherent spectral response and enhanced single- and
two-particle scattering that in turn controls $T_{c}$.  Using constrained RPA, we show further that FeSe in monolayer
form (M-FeSe) provides a natural mechanism to enhance $J$.  We explain why M-FeSe/STO has a high $T_{c}$, whereas M-FeSe
in isolation should not.

Our study opens a paradigm for a unified understanding what controls $T_{c}$ in bulk, layers, and interfaces of Hund's
metals by hole pocket and electron screening cloud engineering.

\end{abstract}

\maketitle


Iron-pnictogen and iron-chalcogen based superconductors (IBS) are classed as Hund's metals, meaning correlations mostly
originate from the Hund's exchange parameter \emph{J}.  In recent years a consensus has evolved that strong Hund's
correlations drive the ubiquitous bad metallicity observed in their phase diagrams~\cite{medici2013,yin1}.  Such
metals have a multiplicity of bands near the Fermi level $E_{F}$; in particular
FeSe~\cite{lanata2013,yin2011,kostin2018} has all five Fe \emph{d} states active there.  Correlations are observed to be
highly orbital-selective~\cite{shen,kostin2018} (a signature of ``Hundness'') with $d_{xy}$ the most strongly correlated
orbital.  However, very little is known whether Hund's correlation can generate glue for superconducting pairing and
control $T_{c}$.

$T_{c}$ is a mere 9\,K in bulk, but it has been observed to increase to $\sim$75\,K when grown as a monolayer on
SrTiO$_{3}$\cite{qing} (M-FeSe/STO), and 109\,K on doped SrTiO$_{3}$~\cite{ge2014}.  Thus while ``Hundness'' has been
found to be important in controlling the single- and two-particle spectral properties of bulk FeSe, the multiplicity of
factors (orbital character, spin-orbit coupling, shape of Fermi surface and dispersion of states around it, differences
in susceptibilities, nematicity, electron-phonon interaction) obfuscate to what extent Hundness, or other factors, drive
superconductivity, and whether `Hundness' can at all explain the jump in $T_{c}$ going from bulk to M-FeSe/STO.

In this work we calculate the superconducting instability using a new high fidelity, \emph{ab initio} approach~\cite{nickel,questaal_paper}.  For the
one-particle Green's function it combines the quasiparticle self consistent \emph{GW} (QS\emph{GW})
approximation~\cite{kotani} with CTQMC solver~\cite{hauleqmc,gull} based dynamical mean field theory (DMFT)~\cite{georges1996}.  This
framework~\cite{prx,Baldini} is extended by computing the local vertex from the two-particle Green's
function by DMFT~\cite{hyowon_thesis,yin}, which is combined with nonlocal bubble diagrams to construct a Bethe-Salpeter equation~\cite{swag19,prl20,boehnke}.  The latter is
solved to yield the essential two-particle spin and charge susceptibilities $\chi^{d}$ and $\chi^{m}$ --- physical
observables which provide an important benchmark.  Moreover they supply ingredients needed for the Eliashberg equation,
which yields eigenvalues and eigenfunctions that describe instabilities to superconductivity.  We will denote
QS\emph{GW}\textsuperscript{\footnotesize{++}} as a shorthand for the four-tier QS\emph{GW}+DMFT+BSE+Eliashberg theory.
The numerical implementation is discussed in Pashov et al.~\cite{questaal_paper} .

QS\emph{GW}\textsuperscript{\footnotesize{++}} has high fidelity because QS\emph{GW} captures non-local dynamic
correlation particularly well in the charge channel~\cite{tomc, questaal_paper}, but cannot adequately capture effects
of spin fluctuations.  DMFT does an excellent job at the latter, which are strong but mostly controlled by a local
effective interaction given by $U$ and $J$.  These are calculated within the constrained RPA~\cite{aryasetiawan} from the QS\emph{GW}
Hamiltonian using an approach~\cite{questaal_paper} similar to that of Ref.~\cite{christoph}.  That it can well describe
superconductivity in a parameter-free manner has now been established in several Hund's materials~\cite{swag19,prl20}.
In FeSe, we have also shown\cite{nemat} that it reproduces the main features of neutron structure
factor~\cite{adroja,boothroyd}.

To isolate the effect of Hundness we make excursions about the \emph{ab initio} reference point, by treating $J$ as a
free parameter.  One of our primary conclusions is that intense and broad low energy spin fluctuations in the vicinity
of an antiferromagnetic ordering vector is the primary glue for pairing and the controlling element for $T_{c}$ in
several variants of FeSe (both bulk and layered); and that this in turn is controlled by \emph{J}.  We further show,
using constrained RPA, that $J$ can be tuned by varying the screening through, e.g. changes in the geometry such as the
change from bulk to monolayer.  We also consider excursions in Fe-Se bond length, and can find correlation can be
enhanced or suppressed by changes in it.  For Hund's coupling to be effective in driving $T_{c}$ it needs a certain
`universal' band feature: the Fe $d_{xy}$ state must be in close proximity to the Fermi energy.

\begin{figure*}[ht!]
	\includegraphics[width=1.0\textwidth]{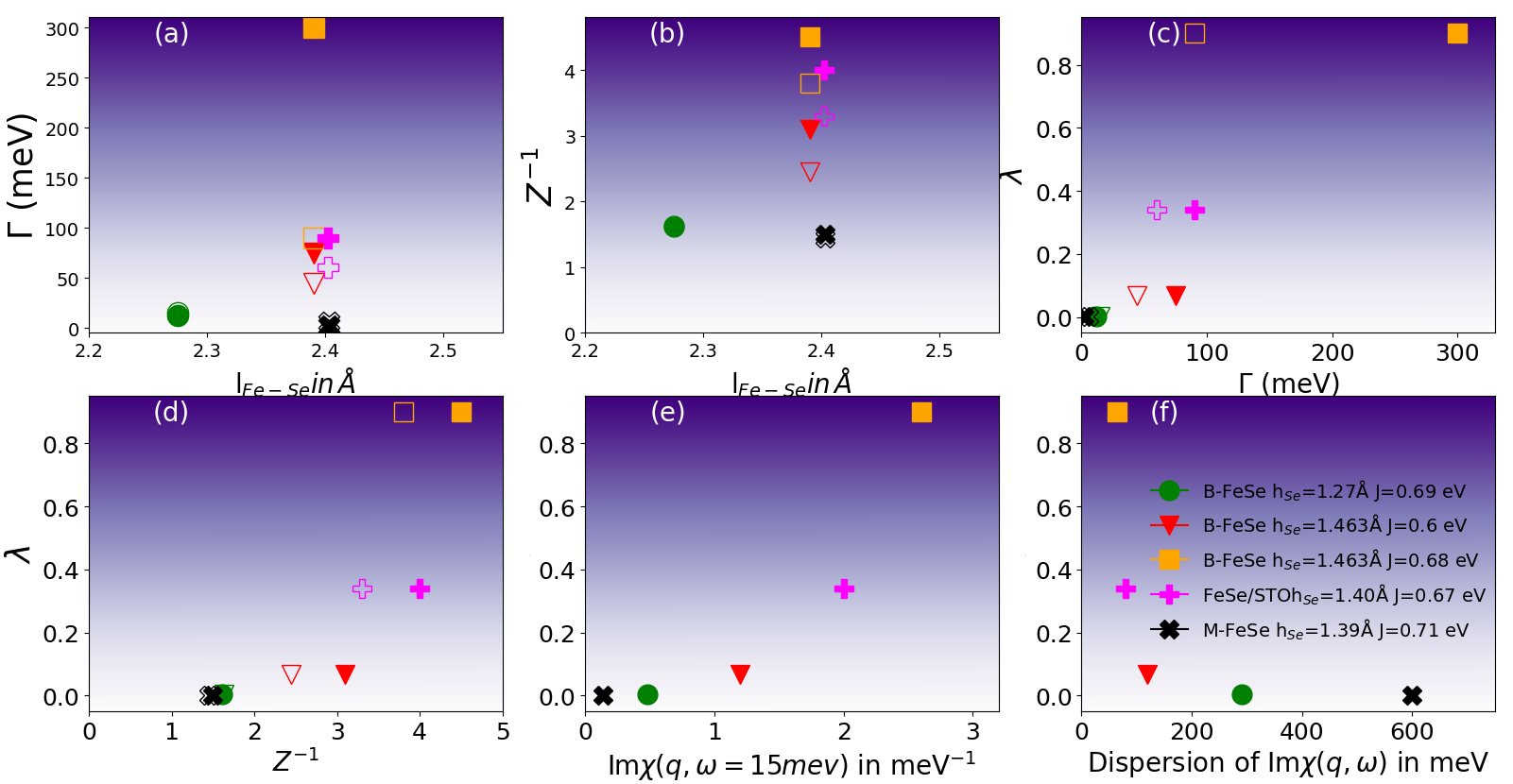}
	\caption{(a,b) Inverse \emph{Z} factors and scattering rates $\Gamma$ for Fe 3$d_{xy}$ (filled symbols) and 3$d_{yz}$ (empty symbols) orbitals
	in three configurations of bulk FeSe ({$\triangledown$,\,\Large$\circ$},\,{\small$\square$})
	$\triangledown$ is the \emph{ab initio} result ($h$=1.463\,\AA, \emph{J}=0.60\,eV); while {\small$\square$}
	changes \emph{J} to 0.68\,eV; {\Large$\circ$} changes \emph{h} and \emph{J} to 1.27\,\AA\ and 0.69\,eV.  Also shown are an
	isolated FeSe monolayer with \emph{h}=1.39\,\AA\ and \emph{J}=0.71\,eV ($\mathbf{\times}$), a monolayer on
	STO with \emph{h}=1.40\,\AA\ and \emph{J}=0.67\,eV ($\mathbf{+}$). Correlation is sensitive to changes in
	$l_{Fe{-}Se}$ and $J$. (\emph{c}--\emph{f}) leading eigenvalue $\lambda$ of the superconducting instability calculated at
	290\,K drawn against various measures of correlation: $\lambda$ is approximately proportional to $\Gamma_{xy}$ and $\Gamma_{yz}$
	($c$), and it is monotonic in $1/Z_{xy}$  and $1/Z_{yz}$ ($d$), and also in the strength of
	Im$\chi[q{=}(1/2,1/2),\omega{=}15\,\mathrm{meV}]$ ($e$) and suppression of the dispersion of the paramagnon branches ($f$). The graded intense purple background separates the most strongly correlated systems with large $\lambda$ from the weakly correlated systems with small  $\lambda$ in weaker purple background.}
	\label{correl}
\end{figure*}


In the crystal, Se sits above and below the Fe plane with height $h{=}1.463$\,\AA\ and bond length
$l_{Fe{-}Se}{=}2.39$\,\AA.  cRPA yields \emph{U}=3.4\,eV and \emph{J}=0.6\,eV from the QS\emph{GW} band structure.  To assess the
role of `Hundness' we also consider how excursions in \emph{J} between 0 and 1\,eV affect the single-particle scattering rate
$\Gamma$, and the inverse \emph{Z} factor, a measure of mass or bandwidth renormalization.
(The SM explains how each are obtained.)
$\Gamma$ is found to be insensitive to \emph{J} for $J<0.4$\,eV, while for $0.4{<}J{<}0.7$\,eV it increases
monotonically for all orbitals.
Thus there is smooth transition from coherence to incoherence with increasing \emph{J}.
$1/Z$ increases from 1.33 at $J{=}0$, reaching a maximum of 4.5 in the $d_{xy}$ channel
at $J{\sim}0.68$\,eV.  Correlation increases for
all states, but $d_{xy}$ is the heaviest and most incoherent, followed by $d_{xz}+d_{yz}$ (see Fig.~\ref{correl}$a$).
For still larger \emph{J}, both $\Gamma_{xy}$ and $1/Z_{xy}$ begin to slowly decrease.
A similar non-monotonic behavior was observed in a previous study of Hund's metals~\cite{dasari}
Similar conclusions were drawn in a recent orbitally-resolved quasi-particle scattering
interference measurement by Kostin et al.~\cite{kostin2018} in the low temperature orthorhombic phase of FeSe, and the orbital
selectivity was emphasized in Ref.~\cite{nica}.

\begin{figure*}[ht!]
	\includegraphics[width=1.00\textwidth]{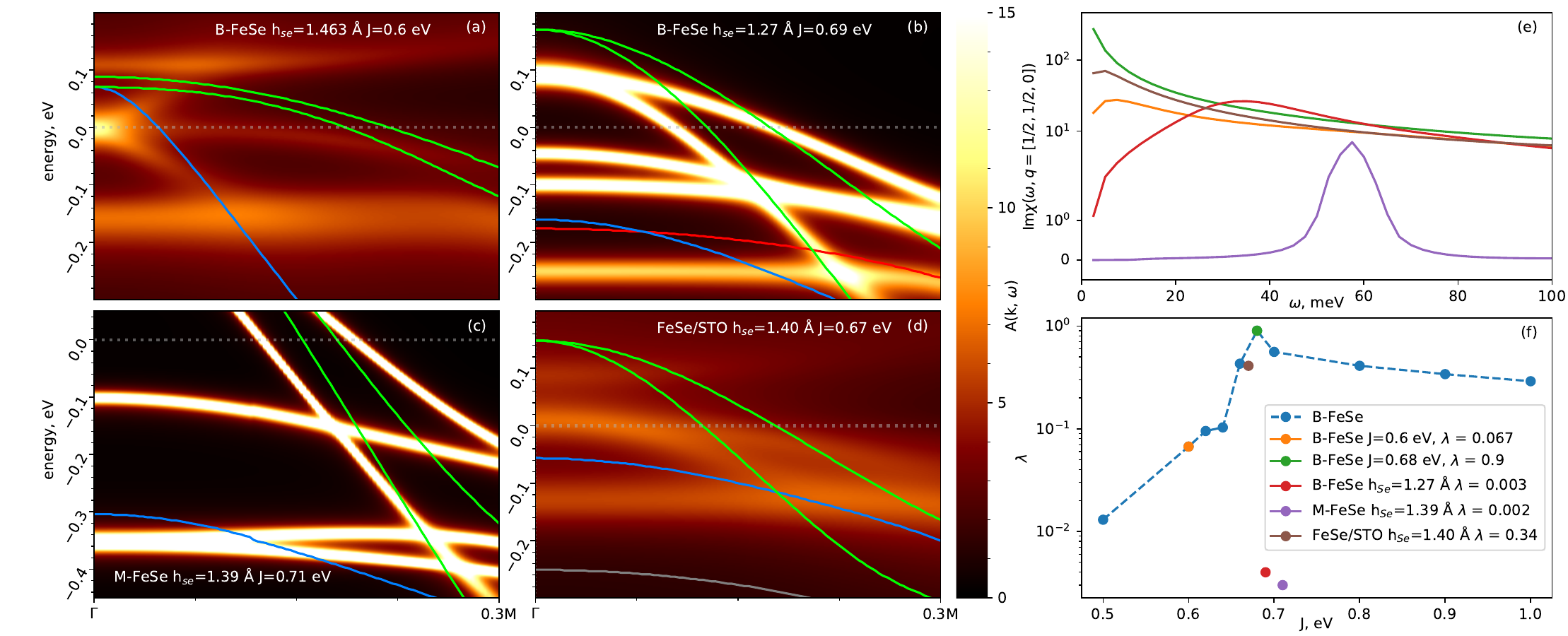}
	\caption{The QS\emph{GW} band structure and QS\emph{GW}+DMFT spectral functions $A(k,\omega)$ are shown on a section
		of the $\Gamma$-M path for: $(a)$ bulk-FeSe ($J$=0.6 eV); $(b)$ bulk FeSe with reduced Se height above the Fe plane
		($h$=1.27\,\AA); $(c)$ M-FeSe, a free standing monolayer of FeSe; (\emph{d}) M-FeSe/STO.  In
		all four panels, the Fe-$d_{xy}$ state calculated by QS\emph{GW} is depicted in blue and the Fermi energy $E_{F}$ is at 0.
		Note the strongly marked incoherence in $(a)$ and $(d)$.
		In all cases DMFT narrows the width of $d_{xy}$ relative to QS\emph{GW} as is typical of narrow-band $d$ systems
		\cite{Tomczak12,nickel}, but incoherence is highly sensitive to the position of $d_{xy}$.
		In (\emph{a}) and (\emph{d}) $d_{xy}$ is proximate to $E_{F}$ and a high degree of incoherence is present.
		while in (\emph{b}) and (\emph{c}) $d_{xy}$ is pushed far below $E_{F}$: and the system has properties similar to a normal Fermi liquid. Panel (\emph{e}) shows the imaginary part of the spin susceptibility $\chi(\omega)$, at the
		AFM nesting vector $\mathbf{q}^\mathrm{AFM}$=(1/2, 1/2, 0)\,$2\pi/a$ for the four geometries (\emph{a}):orange,
		(\emph{b}):red, (\emph{c}):purple, (\emph{d}):brown.  Also shown in green is Im$\chi(\omega)$ for bulk FeSe with $J$=0.68--the
		highest $T_{c}$ found among parameterised hamiltonians.
		The more intense Im\,$\chi(\omega{\rightarrow}0)$ is, the larger the superconducting instability.
		Panel (\emph{f}) shows how the leading eigenvalue $\lambda$ of the linearized
		Eliashberg equation treating $J$ as a free parameter (blue circles).  The extreme sensitivity to $J$ is apparent.  Also
		shown are $\lambda$ for the four \emph{ab initio} calculations (\emph{a-d}), using the colour scheme in panel \emph{e}.
		For ($a$) B-FeSe and (\emph{d}), FeSe/STO, $d_{xy}$ falls near $E_{F}$ and $\lambda$ approximately coincides with the blue line;
		(\emph{b}) and (\emph{c}) do not.}
	\label{summary}
\end{figure*}


Small increases in \emph{J} induce remarkable changes in the transverse spin susceptibility $\chi(q,\omega)$.  The peak
near the antiferromagnetic nesting vector $(1/2,1/2)$ (in 2-Fe unit cell) and $q_{z}$=0 becomes markedly more intense as shown in
Fig.~\ref{correl}$e$ and Fig.~\ref{summary}\emph{e}.  (See SM Fig.\,3 for Im\,$\chi(q,\omega)$ along the
$(0,0)\to(1/2,0)\to(1/2,1/2)\to(0,0)$ lines in the Fe$_{2}$Se$_{2}$ unit cell.)  The energy
dispersion in Im$\chi$ also becomes strongly compressed (Fig.~\ref{correl}$f$).  (Elsewhere~\cite{nemat} we perform a
rigorous benchmarking of $\chi(q,\omega)$ against inelastic neutron scattering measurements~\cite{adroja,lynn2009, boothroyd} where we show that our calculations reproduce all intricate structures of $\chi(q,\omega)$ for all energies and momenta.)
Resolving Im\,$\chi$ into orbital channels, $d_{xy}$ is seen to be the leading component.  Along $(1/2,0){\to}(1/2,1/2)$
it contributes about 50\% of the total with $d_{z^2}$, $d_{x^2-y^2}$ and $d_{xz,yz}$ combining to contribute the rest.

How do these striking changes in Im\,$\chi$ correlate with superconductivity?  We compute the full two particle
scattering amplitude in the particle-particle channel within our DMFT framework, and solve Eliashberg equations in the
BCS low energy approximation~\cite{prl20,swag19,yin,hyowon_thesis}.  Resolving the eigenfunctions of the gap equation
into inter- and intra-orbital channels, two dominant eigenvalues $\lambda$ are found.  Both of them increase with
\emph{J} up to the point of maximum intensity in $\chi$ ($J$=0.68\,eV) and then begin to decrease, as shown in
Fig.~\ref{summary}\emph{f}.  The corresponding eigenfunctions have cos($k_{x}$)+cos($k_{y}$) (leading eigenvalue) and
cos($k_{x}$)-cos($k_{y}$) structures (second eigenvalue)~\cite{hirsh}.  Calculations show that these instabilities reside primarily
in the intra-orbital $d_{xy}$-$d_{xy}$ channel and the inter-orbital components are negligible.  In the bulk crystal,
varying \emph{J} from the \emph{ab initio} value ($\lambda$=0.067 at J=0.60\,eV), we find $\lambda$ reaches its maximum
0.9 at the point where the spin susceptibility is most intense, \emph{J}=0.68\,eV.  We attribute the decrease for
$J{>}0.68$\,eV to the softening of electron masses and loss of spin fluctuations at $q=(1/2,1/2)$ (see SM Fig.\,2 for
$\left<M^2\right>^{1/2}$).


We next consider how parametric changes in the one-body Hamiltonian alter the spectral function, $\chi$ and $T_{c}$.  We
first vary the Fe-Se bond length $l_{Fe{-}Se}$. When it is reduced from its experimental value of 2.39\,\AA, the
$d_{xy}$ band initially near $E_{F}$ at $\Gamma$, gets pushed down well below $E_{F}$.  It is particularly easy to see
at the QS\emph{GW} level (blue band in Fig.~\ref{summary}\emph{b}), reaching about $E_{F}{-}160$\,meV when
{$h$=1.27} or $l_{Fe-Se}$=2.275\,\AA.  From c-RPA we compute the corresponding \emph{U} and \emph{J} as
3.9\,eV and 0.69\,eV respectively.  A similar shift is found in the spectral function calculated by QS\emph{GW}+DMFT
(Fig.~\ref{summary}\emph{b}).  Further, quasi-particles become more coherent: orbital-selective $1/Z$ range between
1.6 to 1.3 and scattering rates become small ({\Large$\circ$}, Fig.~\ref{correl}).  The system behaves as an itinerant
metal, and the peak in Im$\chi(q,\omega)$ at $(1/2,1/2)$ shifts to higher $\omega$ and becomes gapped.  It
also becomes very weak and broad (Fig.~\ref{summary}\emph{e}).  The leading $\lambda$ of the Eliashberg equations
become negligibly small (see Fig.~\ref{summary}\emph{f} and Fig.~\ref{correl}\emph{e}), suggesting extremely weak or
no superconducting instability.  $\lambda$ also becomes completely insensitive to $J$. A similar observation was made in our
recent work on LaFe$_{2}$As$_{2}$~\cite{prl20}, where the collapsed tetragonal phase with lesser $l_{Fe-As}$ in
comparison to its uncollapsed phase, loses superconductivity~\cite{acs} as bands become itinerant due to
increased Fe-Fe hopping mediated via As.


We next turn to a free standing monolayer of FeSe, M-FeSe.  For this study we take the structural inputs from recent
work by Mondal et al.~\cite{mondal2017} which finds the minimum-energy value for $h$ to be 1.39\,\AA\ within a combined
DFT+DMFT framework, while the lattice parameter \emph{a} is somewhat larger, close to that of SrTiO$_{3}$.  (Structural
parameters are given in the SM for all the systems in this study.) As a benchmark, the same work found the equilibrium
$h$ to be close to the measured value in bulk crystalline FeSe~\cite{pascut-force}.  DFT has a tendency to underestimate
$h$; the error is not easily fixed by other kinds of density-functionals.  However, DFT does predict a similar change in
${\delta}h$ between bulk and monolayer which gives us some confidence in the value of $h$.

At the QS\emph{GW} level, the $d_{xy}$ band is pushed to $E_{F}{-}300$\,meV on the $\Gamma$--M line (band structure in
Fig.~\ref{summary}$c$ and SM, bottom panel).  c-RPA calculations yield $U$=4.3\,eV and $J$=0.71\,eV, the increase
arising from reduced screening.  $1/Z$ is found from QS\emph{GW}+DMFT to be 1.5, 1.35, 1.3, 1.25 on $xy$,
${z^2}$, ${yz{+}xz}$, ${x^2{-}y^2}$ respectively (Fig.~\ref{correl}$b$); also a negligible scattering rate is
found (Fig.~\ref{correl}$a$).  As a further indicator of a good metal, Im\,$\chi(q,\omega)$ shows negligible spin
fluctuations in the $d_{xy}$ channel; and at ${q}=(1/2,1/2)$ spin excitations are gapped and vanishingly small
(Fig.~\ref{summary}$e$). The superconducting instability is almost entirely suppressed (see
Fig.~\ref{summary}\emph{f} and Fig.~\ref{correl}\emph{f}).  It is noteworthy that the reduction in electronic
screening reflects in a marked increment in \emph{J}.  Unfortunately, this beneficial effect is more than
counterbalanced by the fact that $d_{xy}$ is pushed far below $E_{F}$ on the scale of magnetic excitation energies.


How does the effect of a SrTiO$_{3}$ substrate modify superconductivity in M-FeSe?  M-FeSe/STO is a subject of intense
debate, since as noted above, $T_{c}$ of M-FeSe/STO has been measured to be an order of magnitude higher than bulk FeSe.
Many explanations have been put forward, e.g. that superconductivity is boosted by large electron-phonon
coupling~\cite{lee2012,xing2014} as SrTiO$_{3}$ is close to a ferroelectric instability (for a contrary view
see~\cite{sawatzky2018}), but the simplest explanation is that SrTiO$_{3}$ modifies M-FeSe to restore $d_{xy}$
to be proximate to $E_{F}$.  M-FeSe/STO is a partially formed Schottky barrier; we can expect the Fermi level to sit in
the SrTiO$_{3}$ bandgap.  SrTiO$_{3}$ modifies M-FeSe in two important ways: the interfacial dipole controls the
Schottky barrier height and changes the electron count in M-FeSe; also the STO (especially the O-$p$-derived bonding
states) couple to the Fe-$d$ in an orbital-selective manner.  Both effects are accurately incorporated by a direct
QS\emph{GW} calculation of M-FeSe/STO.  We consider 5-ML slab of SrTiO$_{3}$ terminated on the Sr side by M-FeSe.  The
structure is relaxed with DFT, subject to the constraint that $h$ is fixed to 1.40\,\AA, as predicted in
Ref.~\cite{mondal2017}. This value is close to $h$=1.39\,\AA\ ($l_{Fe{-}Se}$=2.403\,\AA) found for free-standing
M-FeSe.  Its value is critical, as we have seen in the bulk case, and we cannot rely on DFT for it.

QS\emph{GW} bands and QS\emph{GW}+DMFT spectral functions are shown in Fig.~\ref{summary}$d$.  Also the SM,
Fig.~1($a,e,f$), compares QS\emph{GW} states near $E_{F}$ for the three cases, bulk, M-FeSe and M-FeSe/STO, panels
Fig.~\ref{summary}($a,c,d$), with $h$=1.463, 1.39, 1.40\,\AA, and $a$=3.779, 3.905, and 3.905\,\AA, respectively.
In the last case the SrTiO$_{3}$ states do not appear because they are far from $E_{F}$ (the Schottky barrier height is
calculated to be $\sim$2\,eV, and the direct bandgap $\sim$4\,eV).  As noted above, in M-FeSe $d_{xy}$ is split from the
other Fe-$d$ states and pushed well below $E_{F}$.  This occurs in part from the increase in lattice constant $a$.
However in M-FeSe/STO with the same $a$, this splitting is much reduced (SM Fig.~1$f$). $d_{xy}$ moves to a position
slightly below $E_{F}$; 50 meV in QS\emph{GW} and 100 meV in QS\emph{GW}+DMFT (see Fig.~\ref{summary}$d$) which is fully
consistent with ARPES studies~\cite{ml-arpes}. Also its bandwidth is slightly reduced relative to bulk, in keeping
with a stretched $a$. While our results are remarkable improvement in comparison to the DFT+DMFT (SM) in placing the d$_{xy}$ orbital, we fail to suppress the hole pockets of d$_{xz,yz}$ nature.

A c-RPA calculation for M-FeSe/STO yields $U$=3.8\,eV and $J$=0.67\,eV.  Using QS\emph{GW}+DMFT, we extract the orbital
dependent $1/Z$ (4, 3.3, 3.2 2.7 on $xy$, ${yz{+}xz}$, ${z^2}$, ${x^2{-}y^2}$) and $\Gamma$, which show significantly
enhanced incoherence in the quasiparticle spectrum relative to M-FeSe (see Fig.~\ref{correl}(a,b)).  Further, the
dispersion in Im\,$\chi(q,\omega)$ is significantly narrower than the bulk, and the intensity is spread over momenta in
the region around $(1/2,1/2)$ (see Fig.~\ref{summary}$e$). These signatures suggest that M-FeSe/STO is more correlated
than the both the bulk and free standing monolayer.  The leading instability from the Eliashberg equation survives but the second ($d_{x^2-y^2}$) instability gets suppressed.
Experimentally, the superconducting gap of FeSe/STO show maxima on the bands with $d_{xy}$ character~\cite{ml-gap}. Also
in surface doped bulk FeSe where superconductivity enhances, the appearance of the $d_{xy}$ electron pocket on the Fermi
surface coincides with the beginning of the second superconducting dome with higher T$_{c}$~\cite{ye2015}. The leading
$\lambda$ for superconducting instability is 0.34, which is five times larger that the estimate for bulk FeSe. Our observations strongly contrast the recently published results from Eliashberg theory where it is claimed that spin fluctuations can, at most, account for only two fold increment~\cite{oppeneer} in T$_{c}$. We observe that suppression of the d$_{xz,yz}$ contributions from the hole pockets lead to only 6\% reduction in $\lambda$. Our estimate of five-fold increment is lower than the eight-fold increment in $T_{c}$=75\,K for M-FeSe/STO, but there can be many reasons for this.
The comparison to experiment is not direct: experimentally M-FeSe/STO has a buffer layer between M-FeSe and STO, which
our calculation omits.  Also the calculated $T_{c}$ is extremely sensitive to \emph{J} and $h$.  Both are theoretically
calculated; moreover $h$ is assumed to be the same for the Se planes above and below Fe, while there should be small
differences.  Finally the interface can have several other effects we omitted such as an enhanced phonon contribution to
$T_{c}$. 


To summarize, a unified picture of the origins of superconductivity in FeSe emerges from evidence drawn from several
parametric studies of FeSe around a high-fidelity \emph{ab initio} theory.  Superconducting glue mainly originates from
low-energy spin fluctuations concentrated in a region near the antiferromagnetic ordering vector.  The instability and
the single- and two-particle correlations characterized by the band renormalizations $1/Z$, scattering rate $\Gamma$ and
Im$\chi(q,\omega)$ are all closely linked.  The Fe $d_{xy}$ orbital is the most strongly correlated and contributes
maximally to the pairing glue as long as it is auspiciously near the Fermi level and further, that the Hund's
\emph{J} is sufficiently large to induce a high degree of `bad metallic' behavior.  
Further, the superconducting instability is found to lie predominately in the $d_{xy}$ channel.

Hund's coupling can be used as a parameter to tune $T_{c}$, raising the possibility to enhance $\lambda$ by nearly 15
times. A possible way to control \emph{J} in real materials is to change the electron screening cloud.  We presented two
instances where a structural modification induced a change in $J$ (M-FeSe and M-FeSe/STO).  Controlling number of layers,
applying pressure to tune Fe-chalcogenide bond length, doping and intercalation are some other possibilities. At the
same time it is important to realize it is necessary to control both the screening and specific features of the
single-particle spectrum.  Our calculations show that one promising directions for reaching an optimized $T_{c}$ appears
to be controlling number of layers and interfaces to simultaneously satisfy both conditions: lesser electron screening
leading to a larger Hund's correlation and larger $d_{xy}$ contribution to the Fermiology.

This work was supported by the Simons Many-Electron Collaboration.  We acknowledge SuperMUC-NG at GCS@LRZ, Germany,  Cambridge Tier-2 system operated by the University of Cambridge Research Computing Service  funded by EPSRC Tier-2 capital grant EP/P020259/1.

\pagebreak
\section{Supplemental material}
\begin{widetext}

Here we provide structural parameters used for simulating different systems, the correlation parameters calculated using QS\emph{GW}+contrained-RPA in table~\ref{tab1}.
In table~\ref{tab2} we provide the superconducting eigenvalues computed for different systems for different correlation parameters.
In table~\ref{tab3} we provide orbitally resolved m$_{DMFT}$/m$_{QSGW}$ and scattering rates $\Gamma$ for all compounds. We also show the Fermi surfaces computed from different methods in Fig.~\ref{gap} for both bulk and layered variants of FeSe.

\begin{figure}[ht!]
	\includegraphics[width=0.65\textwidth]{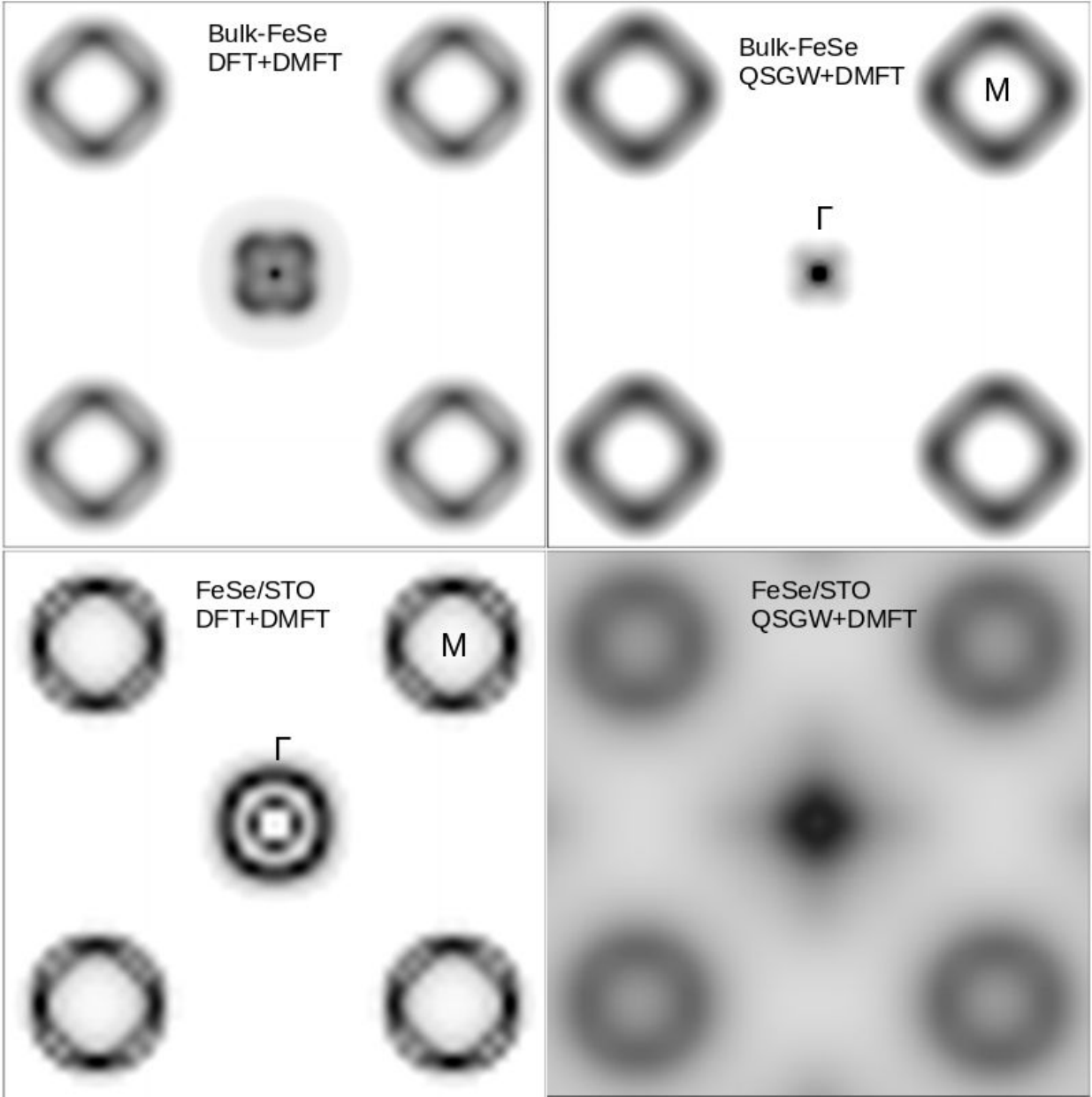}
	\caption{Fermi surfaces are shown in the $\Gamma$-M  (z=0) plane for both bulk FeSe and monolayer FeSe/STO from DFT+DMFT, and QS\emph{GW+DMFT}. In each case the the U and J are used from respective C-RPA calculations. }
	\label{gap}
\end{figure}

\begin{table}[h]
	\begin{center}
		\footnotesize
		\begin{tabular}{ccccccccccc}
			\hline
			variants & \emph{a}\,(\AA) & \emph{c}\,(\AA) & \emph{U}\,(eV)  & \emph{J}\,(eV)  \\
			\hline
			
			B-FeSe ($h_{Se}$=1.463 A$^{0}$)~\cite{kumar}   & 3.779 & 5.5111 & 3.4 & 0.60 \\
			B-FeSe ($h_{Se}$=1.463 A$^{0}$. $\Delta$=0.15)  & do & do & 3.5 & 0.64 \\
			B-FeSe ($h_{Se}$=1.463 A$^{0}$, $\Delta$=0.21)   & do & do & 3.52 & 0.66 \\
			
			B-FeSe ($h_{Se}$=1.27 A$^{0}$) & do & do & 3.9 & 0.69 \\
			
			M-FeSe ($h_{Se}$=1.39 A$^{0}$)~\cite{mondal2017} & 3.905 & & 4.3 & 0.7 \\
			M-FeSe/STO ($h_{Se}$=1.4 A$^{0}$)~\cite{mondal2017} & 3.905 & & 3.8 & 0.67 \\

		\end{tabular}
	\end{center}
	\caption{Structural parameters, chalcogen height $h_{Se}$ and computed \emph{U} and \emph{J} for the
		correlated many body Hamiltonian from our QS\emph{GW}+c-RPA implementation. References indicate where the
		structural inputs were taken. $\Delta$ is the measure for uniform electron doping.}
	\label{tab1}
\end{table}

\begin{table}[h]
	\begin{center}
		\footnotesize
		\begin{tabular}{ccccccccccc}
			\hline
			variants & $J$=0 & 0.5 & 0.6 & 0.62 & 0.64 & 0.66 & 0.68 & 0.7 & 0.8 & 1.0 \\
			\hline
			B-FeSe ($h_{Se}$=1.463 A$^{0}$) & 0.003 & 0.013  & 0.067 & 0.095 & 0.103 & 0.43 & 0.9 & 0.56 & 0.41 & 0.29 \\
			
			B-FeSe ($h_{Se}$= 1.27 A$^{0}$)   &  &   & 0.003 & &  &  & 0.004 \\

			B-FeSe ($\Delta$=0.15)   &  &  & 0.055  &  & 0.09 &  &  & 0.2 &  & \\
			
			B-FeSe ($\Delta$=0.21)   &  &  & 0.04 &  &  & 0.12 &  & 0.18  &  &  \\
			
			M-FeSe ($h_{Se}$=1.39 A$^{0}$)   &  &  &  &  &  &  &  & 0.002 & \\
			M-FeSe/STO ($h_{Se}$=1.4 A$^{0}$)   &  &  &  &  &  &  & 0.34 &  &
			
		\end{tabular}
	\end{center}
	\caption{The leading eigenvalue $\lambda$ from the solution of linearized Eliashberg equations for all
		compounds.  Also shown are results varying \emph{J} (in eV) and  $h_{Se}$ (\AA). For both bulk FeSe and M-FeSe/STO once
		$h_{Se}$ is such that the $d_{xy}$ state contributes to the hole pockets at Fermi level, the superconducting
		eigenvalue becomes sensitive to Hund's coupling \emph{J} and increases with increasing \emph{J}.}
	\label{tab2}
\end{table}

\begin{table}[h]
	\begin{center}
		\footnotesize
		\begin{tabular}{ccccccccccc}
			\hline
			variants & $\Gamma_{x^2-y^2}$ & $Z_{x^2-y^2}$ & $\Gamma_{xz,yz}$ & $Z_{xz,yz}$ & $\Gamma_{z^2}$ & $Z_{z^2}$ & $\Gamma_{xy}$ & $Z_{xy}$\\
			\hline
			B-FeSe (h$_{Se}$=1.27\,\AA) & 18 & 0.74 & 22 & 0.69 & 20 & 0.68 & 34 & 0.61 \\
			
			B-FeSe ($\Delta$=0.15)   & 26 & 0.42 & 32 & 0.35 & 2 & 0.38 & 21 & 0.32\\
			
			B-FeSe ($\Delta$=0.21) & 18 & 0.5 & 26 & 0.52 & 23 & 0.58 & 42 & 0.46 \\
			
			M-FeSe ($h_{Se}$=1.39\,\AA) & 14 & 0.8 & 17 & 0.76 & 10 & 0.74 & 20 & 0.68 \\
			M-FeSe/STO ($h_{Se}$=1.40\,\AA) & 33 & 0.37 & 60 & 0.3 & 40 & 0.31 & 120 & 0.25 \\
			
		\end{tabular}
	\end{center}
	\caption{Quasi-particle renormalization factor $Z$, single-particle scattering rate $\Gamma$ for bulk FeSe with
		smaller $h_{Se}$, doped FeSe,  and M-FeSe/STO and different choices of $h_{Se}$. $\Gamma$ is in meV.}
	\label{tab3}
\end{table}

For electronic band structure for all materials over the relevant BZ paths, see QS\emph{GW} band structure (see Fig.~\ref{band}).
Parameters extracted from the DMFT self energy $\Sigma(\omega)$ (see Table~\ref{tab3} and Fig.~\ref{coherence}).  The single-site DMFT Im$\Sigma(i\omega)$ is fitted to a fourth order polynomial in
$i\omega$ for low energies (first 6 matsubara points at $\beta$=40 eV$^{-1}$=290 K). The mass enhancement, related to the coefficient ($s_{1}$) of the linear term in the expansion $m_{DMFT}/m_{QSGW} =
1+|s_{1}|$~\cite{millis}, and the intercept $|s_{0}|$=$\Gamma m_{DMFT}/m_{QSGW}$.  m$_{DMFT}$/m$_{QSGW}$ = $Z^{-1}$ is resolved in different intra-orbital channels.

\begin{figure}[ht!]
	\includegraphics[width=0.64\textwidth]{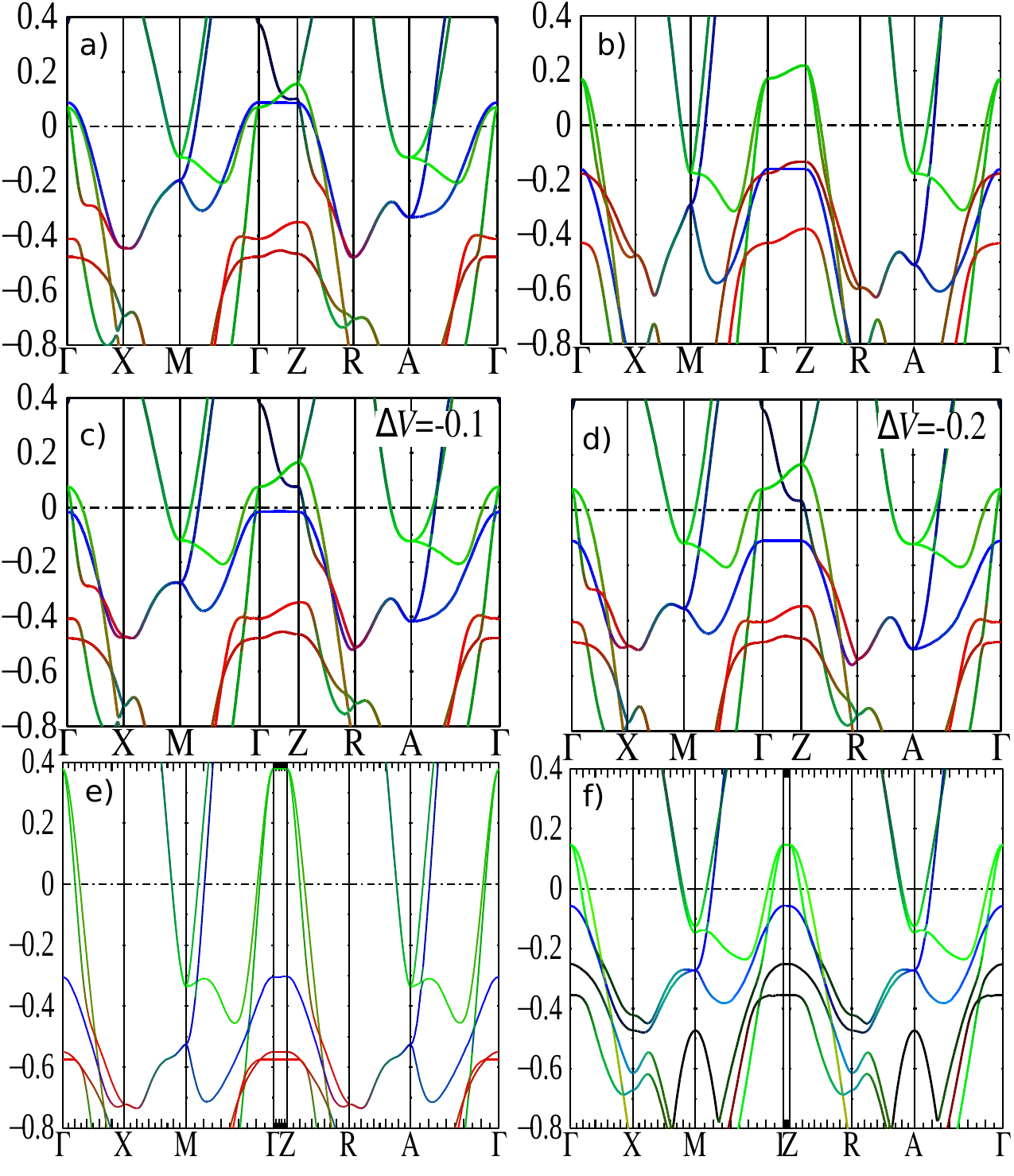}\\
	\caption{Color weighted QS\emph{GW} band structure in the wondow ($-$0.8,\, 0.4)\,eV ($E_{F}{=}0$); (a) is for $h_{Se}$ at its equilibrium value (1.463 $\AA$ in the crystalline case), (b) is for $h_{Se}$= 1.27 $\AA$.  Red, green and
		blue colors show projections onto Fe $e_{g}$, $d_{xz}{+}d_{yz}$, and $d_{xy}$ orbital characters, respectively.
		(c) and (d) show QS\emph{GW} band structure when modified by a combination of $U$ added to the QS\emph{GW}
		Hamiltonian combined with a constant background charged added to preserve the position of $E_{F}$, as discussed in the
		text.  Note the close similarity with the top left panel, except for the shift in the $d_{xy}$ (blue) band. (c): potential shift is $\Delta\,V{=}nU{=}-0.1$\,eV,  (d): potential shift is $\Delta\,V{=}nU{=}-0.2$\,eV, (e) shows a free standing monolayer of M-FeSe and the (f) shows the FeSe/STO. The $d_{xy}$ (blue) band can be found to be pushed below the Fermi energy, more severely in M-FeSe.}
	\label{band}
\end{figure}


\begin{figure*}[ht!]
	\includegraphics[width=0.32\textwidth]{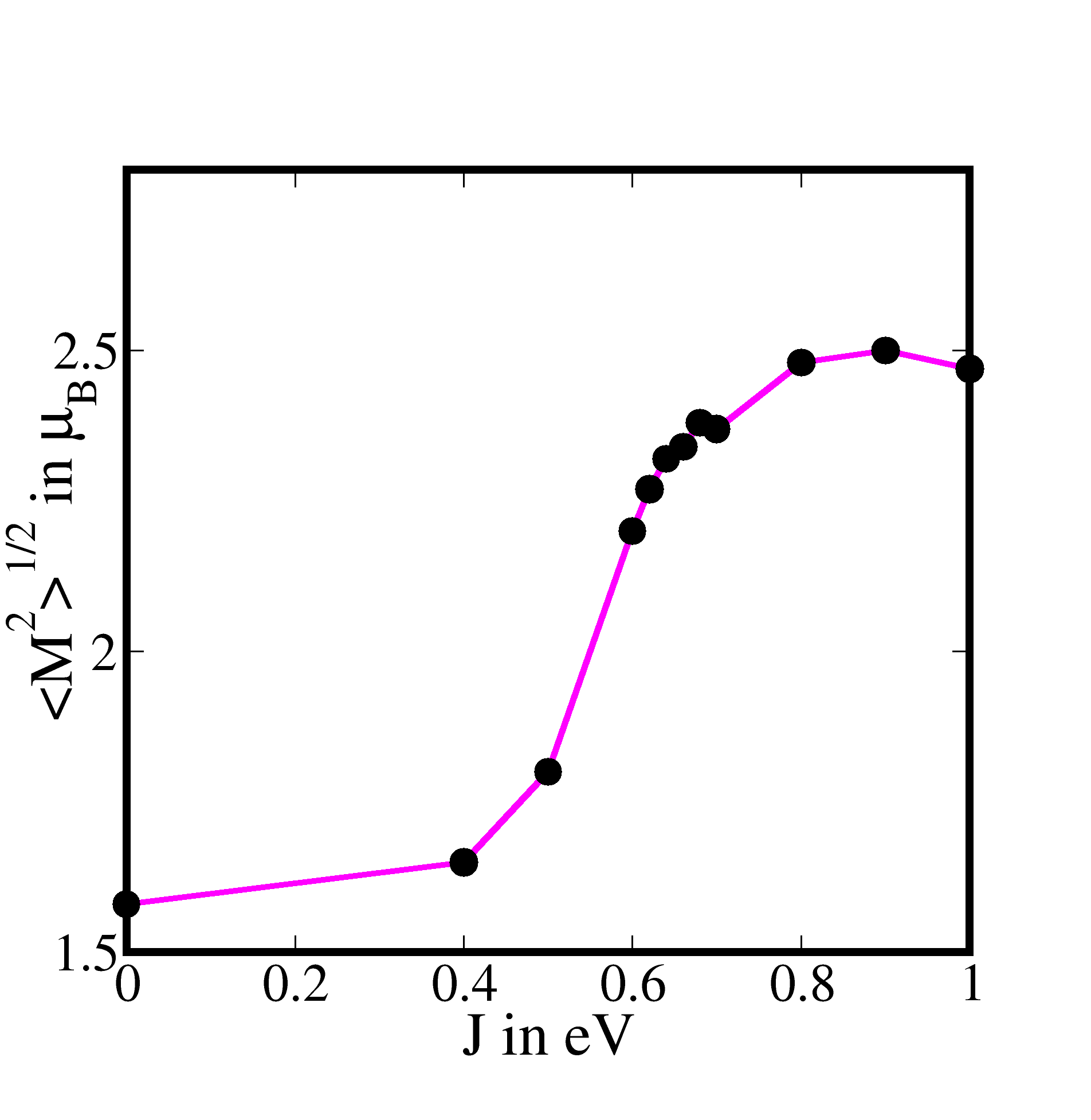}	
	\includegraphics[width=0.32\textwidth]{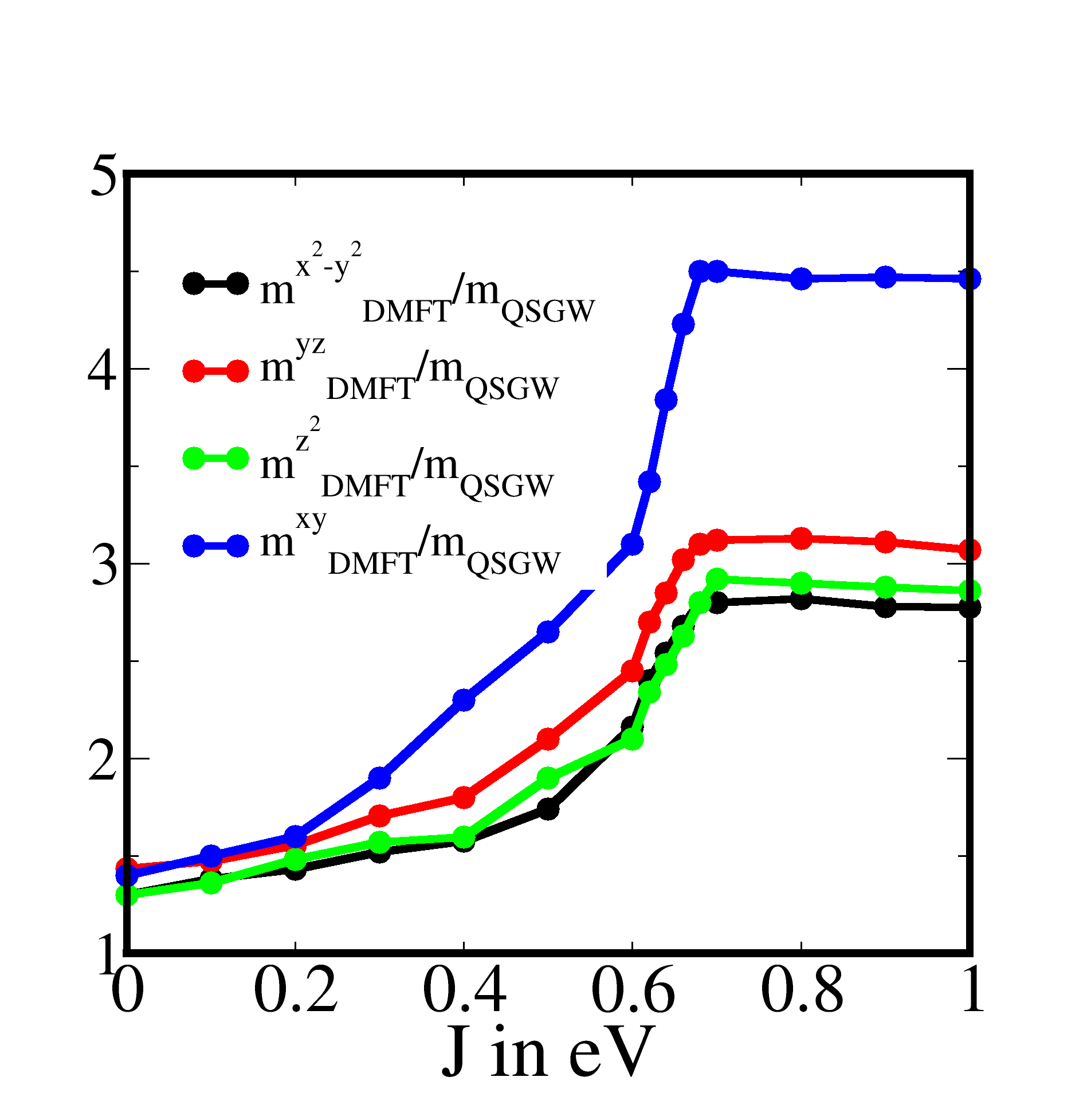}
	\includegraphics[width=0.32\textwidth]{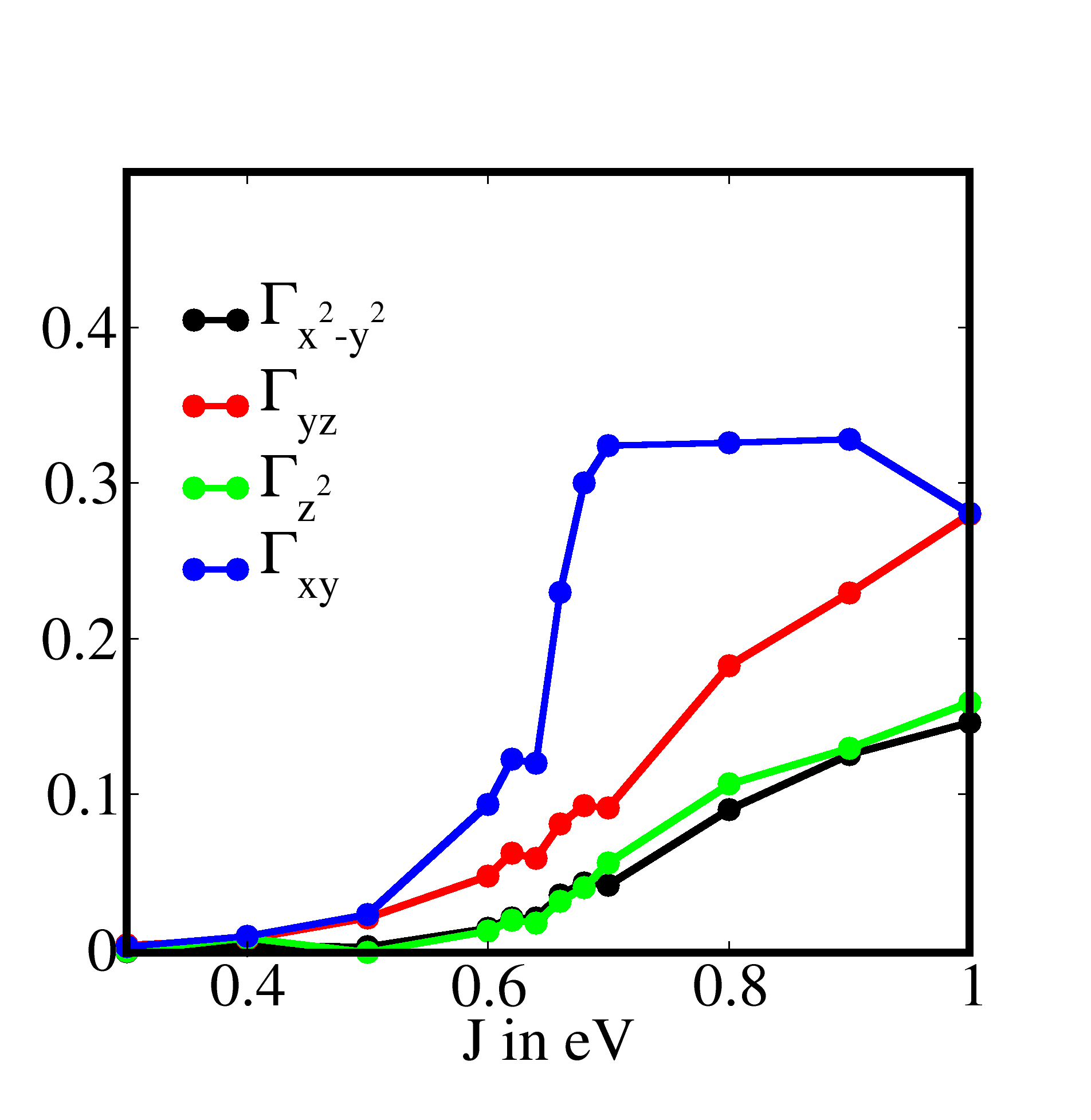}	
	\caption{We compute the net local moment and it's evolution with J. Orbitally resolved single-particle scattering rate ($\Gamma$) and mass enhancement m$_{DMFT}$/m$_{QSGW}$  for Bulk FeSe with varying Hund's coupling strength.}
	\label{coherence}
\end{figure*}

All DMFT and DMFT+BSE results presented in the main text are performed at 290 K.  DMFT is solved for all five Fe-$3d$
orbitals using a Continuous time Quantum Monte Carlo technique (CTQMC)~\cite{hauleqmc} on a rotationally invariant Coulomb interaction.

We also show the single-particle QS\emph{GW}+DMFT spectral functions A(k,$\omega$)~\ref{spectra} and the imaginary part of dynamic and momentum resolved spin susceptibility Im$\chi(q,\omega)$~\ref{chi} for all relevant directions along in the BZ. 

\begin{figure*}[ht!]
	\includegraphics[width=1.0\textwidth]{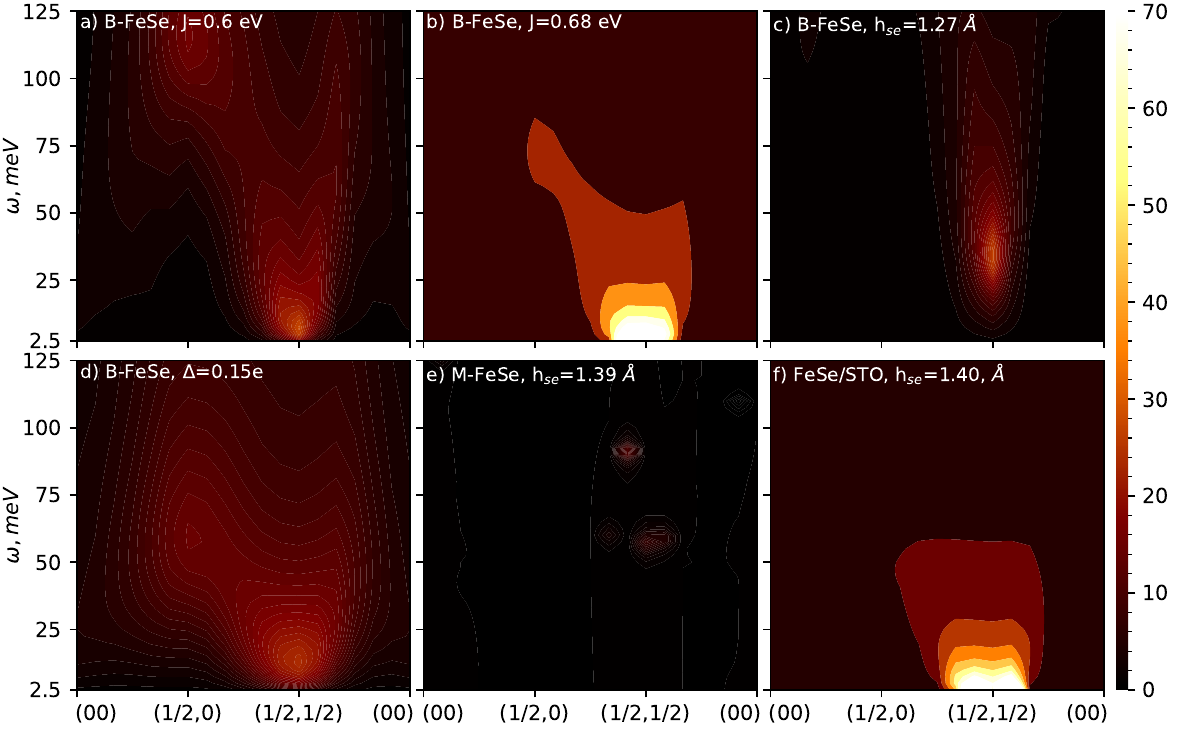}
	\caption{Energy and momentum resolved spin susceptibility Im$\chi(q,\omega)$ shown for (a) bulk FeSe (B-FeSe) (J=0.6
		eV), (b) bulk FeSe with increased Hund's correlation ($J$=0.68 eV), (c) reduced Fe-Se height ($h_{Se}$=1.27 $\AA$),
		(d) 0.15 electron doped bulk FeSe, (e) free standing monolayer of FeSe, M-FeSe~\cite{mondal2017} (f)
		M-FeSe/STO~\cite{mondal2017}. The q-path (H,K,L=0) chosen is along
		(0,0)-($\frac{1}{2},0$)-($\frac{1}{2},\frac{1}{2}$)-(0,0) in the Brillouin zone corresponding to the two Fe-atom unit
		cell. The intensity of the spin fluctuations at ($\frac{1}{2},\frac{1}{2}$) is directly related to the presence of the
		Fe-d$_{xy}$ state at Fermi energy and its incoherence. The more incoherent the A(k,$\omega$) is the more intense is
		the Im\,$\chi(\mathbf{q}=(\frac{1}{2},\frac{1}{2},0),\omega)$.  }
	\label{chi}
\end{figure*}

\begin{figure*}[ht!]
	\includegraphics[width=1.00\textwidth]{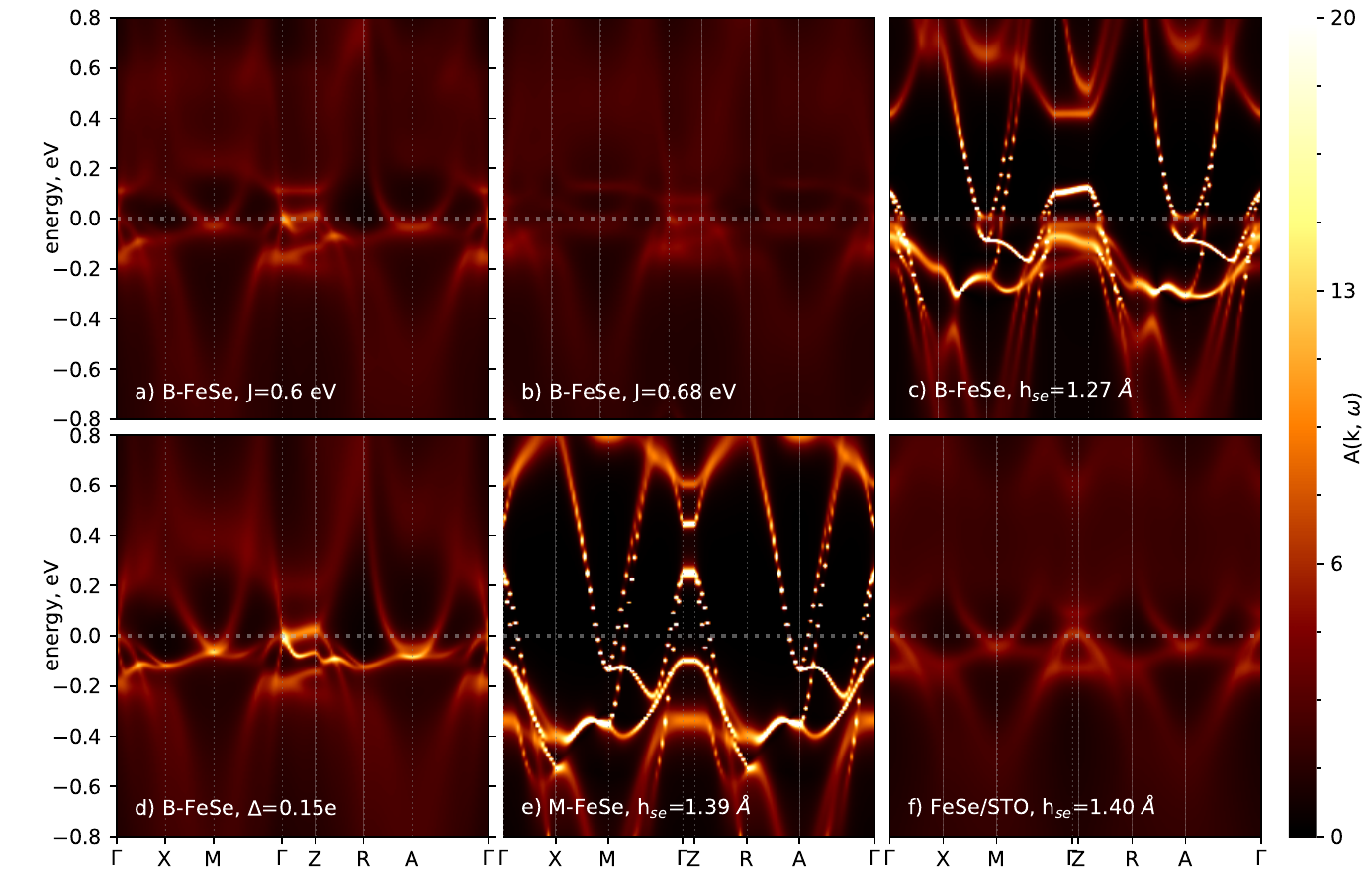}
	\caption{Spectral function $A(k,\omega)$ in FeSe. (\emph{a}) bulk FeSe ($J$=0.6 eV), (\emph{b}) same with Hund's
		coupling increased to 0.68\,eV, (\emph{c}) same with reduced Se height above the Fe plane
		($h_{Se}$=1.27\,\AA). (\emph{d}) 0.15 electron doped bulk FeSe, (\emph{e}) free standing monolayer of FeSe,
		M-FeSe~\cite{mondal2017} (\emph{f}) M-FeSe/STO~\cite{mondal2017}.  The position of the Fe $d_{xy}$ band centre is very
		sensitive to $h_{Se}$, and decreases as $h_{Se}$ decreases.  At values $h_{Se}$ when put $d_{xy}$ is in close
		proximity to $E_{F}$, incoherence in the spectral features are enhanced, and $T_{c}$ increases.
		When $h_{Se}$ is low (panels \emph{c} and \emph{e}) $d_{xy}$ is pushed well below $E_{F}$:
		coherent spectral features result and $T_{c}$ is low.  The incoherence is also highly sensitive to Hund's coupling
		(compare panels (\emph{a}) and (\emph{b})).
	}
	\label{spectra}
\end{figure*}

We also discuss in details the methods implemented for electron doping bulk-FeSe with uniform amount of charge, and the results in subsequent sections.

\subsection{Doped FeSe}
To better establish the role of the $d_{xy}$ orbital in controlling the transition from bad-metal with
superconductivity to a good metal without it, we parameterize the QS\emph{GW} Hamiltonian around its \emph{ab initio}
point in a carefully controlled manner that targets the band center of this orbital keeping everything else nearly
constant.  We add an artificial Hubbard $U$ to the QS\emph{GW} Hamiltonian, on the Fe $3d$ states, using an artificial
(fixed) density matrix $n$ that shifts only the potential on the $d_{xy}$ partial wave.  To compensate for the shift in
$E_{F}$ caused by the shift in $d_{xy}$, we simultaneously add a uniform background charge $Q$.  The net effect is
to shift states of $d_{xy}$ character while leaving all of the other states nearly unchanged.  We performed this
parameterization for two sets of ($n$, $U$, $Q$): the first inducing a shift ${\Delta}V=nU=-0.1$\,eV, compensated by
0.155$e$ background charge, the second shifting $d_{xy}$ by ${\Delta}V=nU=-0.2$\,eV, compensated with 0.211$e$
background charge.  The QP band structures are shown in Fig.~1 of the SM.  They are very similar to the \emph{ab initio}
counterpart, apart from the shift in the $d_{xy}$ band.  For both cases $d_{xy}$ is pushed below $E_{F}$ at $\Gamma$ in
QS\emph{GW}, but DMFT simultaneously renormalizes the position of this band and broadens it, hence $d_{xy}$ still
contributes to low energy scattering.  c-RPA calculations give \emph{J}=0.64 and 0.66\,eV respectively for background
charges $\Delta$=0.15 and 0.21. We find that there low energy spin fluctuations remain around $\mathbf{q}{=}(1/2,1/2)$,
but they are weakly suppressed and diffused relative to the \emph{ab initio} case (compare Fig.~\ref{chi}(\emph{a}) to
Fig.~\ref{chi}(\emph{d})).  This results in $T_{c}$ decreasing from 9\,K to 6\,K.


These parametric studies isolating the role $J$ and the position of $d_{xy}$, clearly establish that for crystalline
FeSe, $T_{c}$ is controlled by these two key parameters.  For $T_{c}$ to be high, the $d_{xy}$ state must be close to
$E_{F}$, possibly crossing it.  Once this favorable band structure emerges, large Hund's coupling can strongly enhance
the single-particle incoherence and subsequently confine spin fluctuations to low energies and enhance the intensity in
the vicinity of the antiferromagnetic $q$ vector.  In bulk FeSe, the position of $d_{xy}$ is favorable, but
$T_{c}$ is only 9\,K because $J$ is sub-optimal.


\end{widetext}


%

\end{document}